\documentclass{article}

\usepackage{arxiv}

\usepackage[utf8]{inputenc} 
\usepackage[T1]{fontenc}    
\usepackage{hyperref}       
\usepackage{url}            
\usepackage{booktabs}       
\usepackage{amsfonts}       
\usepackage{nicefrac}       
\usepackage{microtype}      
\usepackage{graphicx}
\usepackage[numbers]{natbib}
\usepackage{doi}
\usepackage{enumitem}
\usepackage{makecell}
\usepackage{titlesec}
\usepackage{amsmath}
\usepackage{caption}
\usepackage[table]{xcolor}
\usepackage{multicol}
\setlist{leftmargin=3.0mm}

\titlespacing\subsection{0pt}{12pt plus 4pt minus 2pt}{0pt plus 4pt minus 2pt}

\title{Enhancing composition-based materials property prediction by cross-modal knowledge transfer}

\author{
	Ivan Rubtsov \\
	Lomonosov Moscow State University\\
	Moscow 119991, Russia\\
	\And
	Ivan Dudakov \\
	Lomonosov Moscow State University\\
	Moscow 119991, Russia\\
	\And
	\href{https://orcid.org/0009-0007-0718-0824}{\includegraphics[scale=0.07]{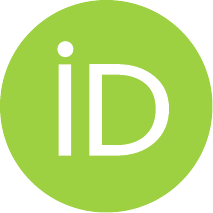}\hspace{1mm}Yuri Kuratov} \\
	Lomonosov Moscow State University\\
	Moscow 119991, Russia\\
    \And
    \href{https://orcid.org/0000-0001-6117-5662}{\includegraphics[scale=0.07]{orcid.pdf}\hspace{1mm}Vadim Korolev}\thanks{\textit{Email address}: \texttt{v.korolev@iai.msu.ru}} \\
	Lomonosov Moscow State University\\
	Moscow 119991, Russia\\    
}

\renewcommand{\shorttitle}{Enhancing composition-based materials property prediction}

\begin{document}
\maketitle

\begin{abstract}
Crystal graph neural networks are widely applicable in modeling experimentally synthesized compounds and hypothetical materials with unknown synthesizability. In contrast, structure-agnostic predictive algorithms allow exploring previously inaccessible domains of chemical space. Here we present a universal approach for enhancing composition-based materials property prediction by means of cross-modal knowledge transfer. Two formulations are proposed: implicit transfer involves pretraining chemical language models on multimodal embeddings, whereas explicit transfer suggests generating crystal structures and implementing structure-aware predictors. The proposed approaches were benchmarked on LLM4Mat-Bench and MatBench tasks, achieving state-of-the-art performance in 25 out of 32 cases. In addition, we demonstrated how another modeling aspect of chemical language models—interpretability—benefits from applying a game-theoretic approach, which is able to incorporate high-order feature interactions.
\end{abstract}

\section{Main}
\label{sec:main}
The duality between material’s constituent parts and its properties, metaphorically referred to as the incarnation of the philosophical duality between body and soul\cite{zunger2018inverse}, is increasingly resolved by means of machine learning approaches\cite{butler2018machine}. Task specificity determines the granularity of materials representation at which prediction model operates. High-throughput computational screening across synthetically accessible compounds is typically carried out using structure-aware models that take into account the crystallographic data. Accuracy of graph neural networks (GNNs) utilized in materials science\cite{reiser2022graph} has been consistently improved by introducing the convolution operation on crystal graphs\cite{xie2018crystal}, learnable bond and global-state embeddings\cite{chen2019graph}, many-body interactions\cite{choudhary2021atomistic}, and neighbor equalization\cite{sole2025cartesian}. Recently, further performance gains have been achieved with multimodal architectures, which incorporate data beyond the spatial arrangement of atoms\cite{moro2025multimodal}. Considering the immense size of chemical space\cite{davies2016computational}, a much more ambitious strategy is exploring compounds that were not studied experimentally before. Transfer from stoichiometric composition to crystalline phase, i.e., crystal structure prediction (CSP)\cite{oganov2019structure}, is a notorious computational task impeding the use of structure-aware models in exploring unknown chemical domains. In this regard, composition-based property predictors have been extensively developed as well\cite{alghadeer2024machine}, starting with classical ML algorithms trained on hand-crafted features\cite{ward2016general,zhuo2018predicting}; the descriptors constructed as analytical expressions deserve special attention\cite{ghiringhelli2015big,ouyang2018sisso}. Lately, deep learning approaches have shifted the focus from manual feature engineering to the development of universal representations that capture domain knowledge inherently. A pioneering model of this kind, ElemNet\cite{jha2018elemnet}, was based on a 17-layered fully-connected architecture, whereas the input vectors consisted of element fractions. Its successors\cite{goodall2020predicting,wang2021compositionally,chen2021atomsets,ihalage2022formula} have been advanced by incorporating pretrained element embeddings\cite{tshitoyan2019unsupervised} and attention mechanisms\cite{vaswani2017attention,velivckovic2018graph} into training. Diverse pretraining strategies, including self-supervised learning, fingerprint learning, and multimodal learning, have been applied to improve the efficacy of the Representation Learning from Stoichiometry (Roost) framework\cite{goodall2020predicting} on downstream tasks. One more step in developing accurate structure-based models has been made by Na\cite{na2025cross}, who proposed the cross-modal transfer learning approach for modeling experimental properties.

\begin{figure}
  \centering
  \includegraphics{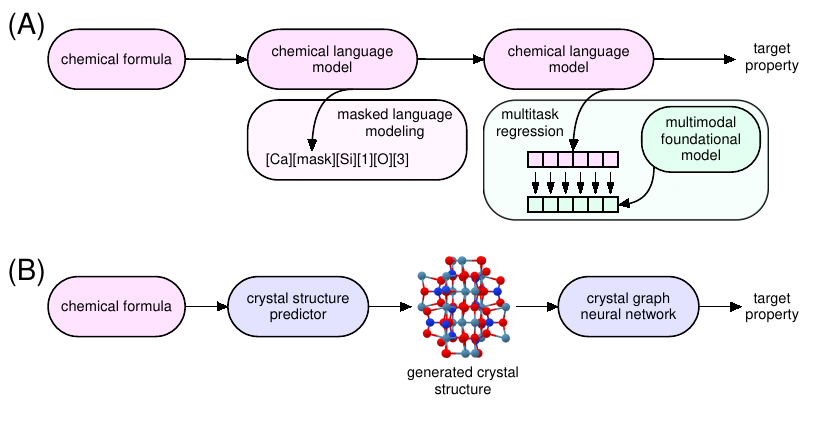}
  \caption{\textbf{A conceptual scheme of cross-modal knowledge transfer formulations.} (A) Implicit knowledge transfer pipeline includes two pretraining phases: masked language modeling (on chemical symbols and stoichiometric coefficients) and multitask regression on multimodal embeddings produced within the foundational model (MultiMat). (B) Explicit knowledge transfer pipeline involves crystal structure prediction in a high-throughput manner and materials property prediction using structure-aware model.}
  \label{fig:fig1}
\end{figure}

The aforementioned deep learning approaches incorporate chemical composition as a set of elemental or atom-wise attributes, whereas the advent of chemical language models (CLMs) reframes the composition-based property prediction as a sequence modeling task. The models originally trained via masked language modeling (MLM) on materials science abstracts\cite{trewartha2022quantifying} and crystal text descriptions\cite{niyongabo2025llm} have been finetuned and evaluated in reproducing diverse quantities\cite{rubungo2025llm4mat}. Here, we take a next step in developing CLMs aimed at composition-based property prediction. In contrast to the previously mentioned MatBERT\cite{trewartha2022quantifying} and LLM-Prop\cite{niyongabo2025llm} models, multiple modalities beyond natural language were incorporated into pretraining (Figure 1A). After the MLM stage, embeddings from CLMs were aligned to those of the foundation model recently presented under the framework of multimodal learning for materials\cite{moro2025multimodal} (MultiMat). We utilized the crystal structure encoder that was contrastively pretrained on four materials modalities: crystal structure, density of electronic states, charge density, and textual description. This approach can be characterized as \textit{implicit} cross-modal Knowledge Transfer (imKT) since the considered property predictors, CLMs, directly operate on modality of interest, i.e., chemical composition. Alternatively, materials property prediction was \textit{explicitly} transferred (exKT) from the compositional to structural domain with the large language model—CrystaLLM\cite{antunes2024crystal}—served as a crystal structure predictor, followed by GNN finetuned on the generated crystals (Figure 1B). The rest of the paper is devoted to the demonstration of how cross-modal knowledge transfer affects predictive performance, taking into account the current state-of-the-art (SOTA) algorithms. In addition, explainability analysis of CLM outputs is presented, considering high-order token interactions.

Table 1 presents a comparison of our best-performing models with SOTA algorithms available in the literature; a comprehensive picture of how accurately all models perform is provided in Tables S1–S5. Considering 20 tasks from the JARVIS-DFT dataset (LLM4Mat-Bench\cite{rubungo2025llm4mat}), substantial improvement (from 4.5 to 39.6\%) in mean absolute error (MAE) was achieved in 18 cases; on average, the MAE decreased by 15.7\%. Knowledge transfer failed to reduce MAE for two tasks: energy above the convex hull and maximal piezoelectric strain coefficient. Prediction of the former property is expectedly complicated by introducing stability-aware weights in the MLM loss function (see details in Supporting Information, Methods section), so CLMs pretrained on multimodal embeddings surpassed models trained via the two-stage procedure (Table S6). By applying the imKT approach, we also significantly improved performance in predicting four band-gap-related tasks from the SNUMAT dataset (LLM4Mat-Bench); on average, the MAE decreased by 15.2\%.

\begin{table}
\caption{\textbf{Predictive performance comparison of transfer knowledge models with existing architectures.} Three separated groups of tasks correspond to JARVIS-DFT, SNUMAT (both are the LLM4Mat-Bench datasets), and MatBench benchmarks. Mean absolute error (MAE) values are provided for the models demonstrating state-of-the-art (SOTA) performance.}
\begin{center}
\footnotesize
\setlength{\tabcolsep}{2.5pt} 
\renewcommand{\arraystretch}{1.2} 
\begin{tabular}{c c c c c c c}
 \hline
 & \multicolumn{2}{c}{SOTA existing models} & & \multicolumn{2}{c}{SOTA presented models} \\
 \makecell{predictive task} & \makecell{architecture} & \makecell{MAE $\downarrow$} & \makecell{performance \\ boost $\uparrow$} & \makecell{MAE $\downarrow$} & \makecell{architecture} \\
 \hline
 FEPA	& MatBERT-109M	& 0.126	& +8.8\%	& 0.11488 ± 0.00018	& imKT@ModernBERT \\
 Band gap (OPT)	& MatBERT-109M	& 0.235	& +15.5\%	& 0.1985 ± 0.0019	& imKT@BERT \\
 Total energy	& MatBERT-109M	& 0.194	& +39.6\%	& 0.1172 ± 0.0005	& imKT@ModernBERT \\
 Ehull	& MatBERT-109M	& 0.096	& –	& 0.1031 ± 0.0009	& imKT@RoFormer \\
 Band gap (MBJ) &	MatBERT-109M	& 0.491	& +23.2\%	& 0.3773 ± 0.0030	& imKT@ModernBERT \\
 Kv	& MatBERT-109M	& 18.498	& +11.6\%	& 16.35 ± 0.24	& imKT@ModernBERT \\
 Gv	& MatBERT-109M	& 14.241	& +10.4\%	& 12.76 ± 0.05	& imKT@ModernBERT \\
 SLME	& MatBERT-109M	& 5.851	& +16.1\%	& 4.911 ± 0.010	& imKT@ModernBERT \\
 Spillage	& MatBERT-109M	& 0.409	& +15.4\%	& 0.3462 ± 0.0029	& imKT@ModernBERT \\
 ${\epsilon}_{x}$ (OPT) & MatBERT-109M	& 32.661	& +25.5\%	& 24.32 ± 0.06	& imKT@ModernBERT \\
 ${\epsilon}$	& Gemma2-9b-it:5S	& 28.228	& +5.8\%	& 26.6 ± 0.4	& imKT@RoFormer \\
 Max. piezo. (${d}_{ij}$)	& Gemma2-9b-it:5S	& 7.973	& –	& 9.67 ± 0.10	& imKT@ModernBERT \\
 Max. piezo. (${e}_{ij}$)	& LLM-Prop-35M	& 0.156	& +4.5\%	& 0.1490 ± 0.0026	& imKT@BERT \\
 Max. EFG	& MatBERT-109M	& 26.621	& +12.5\%	& 23.30 ± 0.11	& imKT@ModernBERT \\
 Exfoliation energy	& MatBERT-109M	& 37.445	& +21.2\%	& 29.5 ± 1.4	& imKT@RoFormer \\
 avg. ${m}_{e}$	& MatBERT-109M	& 0.103	& +18.7\%	& 0.0837 ± 0.0010	& imKT@ModernBERT \\
 n-Seebeck	& MatBERT-109M	& 58.342	& +16.7\%	& 48.6 ± 0.5	& imKT@ModernBERT \\
 n-PF	& MatBERT-109M	& 528.070	& +6.5\%	& 493.7 ± 1.7	& imKT@ModernBERT \\
 p-Seebeck	& MatBERT-109M	& 61.085	& +17.8\%	& 50.22 ± 0.06	& imKT@ModernBERT \\
 p-PF	& LLM-Prop-35M	& 544.737	& +12.2\%	& 478.5 ± 1.4	& imKT@ModernBERT \\
 \hline
 Band gap GGA	& MatBERT-109M	& 0.461	& +19.9\%	& 0.3694 ± 0.0009	& imKT@ModernBERT \\
 Band gap HSE	& MatBERT-109M	& 0.553	& +21.5\%	& 0.4341 ± 0.0027	& imKT@ModernBERT \\
 Band gap GGA optical	& MatBERT-109M	& 0.701	& +10.3\%	& 0.629 ± 0.004	& imKT@ModernBERT \\
 Band gap HSE optical	& MatBERT-109M	& 0.749	& +9.1\%	& 0.6811 ± 0.0015	& imKT@ModernBERT \\
 \hline
 Castelli perovskites	& AtomSets	& 0.082 ± 0.001	& –	& 0.149 ± 0.010	& imKT@RoFormer \\
 Refractive index	& Roost-SSL	& 0.3122 ± 0.0808	& –	& 0.35 ± 0.09	& imKT@RoFormer \\
 log${}_{10}$(shear modulus)	& CrabNet	& 0.092	& +4.8\%	& 0.0876 ± 0.0020	& imKT@ModernBERT \\
 log${}_{10}$(bulk modulus)	& CrabNet	& 0.068	& +1.6\%	& 0.0669 ± 0.0031	& imKT@ModernBERT \\
 Experimental band gap	& CrabNet	& 0.338	& +7.7\%	& 0.312 ± 0.022	& imKT@RoFormer \\
 MP formation energy	& CrabNet	& 77	& –	& 78.9 ± 1.7	& imKT@ModernBERT \\
 MP band gap	& Finder	& 0.231	& –	& 0.253 ± 0.004	& imKT@RoFormer \\
 Phonon peak	& Roost-SSL	& 46.05 ± 4.22	& –	& 54 ± 4	& imKT@RoFormer \\
 \hline
\end{tabular}
\end{center}
\label{table:3}
\end{table}

LLM4Mat-Bench is the largest benchmark for evaluating performance of CLMs in predicting properties of crystalline materials, lacking the assessment of composition-based predictors based on other neural network architectures. To fill this gap, we utilized another well-established suite of predictive tasks, MatBench\cite{dunn2020benchmarking} (Tables 1, S4, S5). The imKT models achieved SOTA results on only three out of eight considered tasks; there is no dominating architecture that provides the best performance on most tasks, as opposed to the LLM4Mat-Bench datasets. An integrative comparison across datasets was done by calculating the weighted average of MAD:MAE ratio, where MAD stands for the mean absolute deviation. The best value of 8.25 was achieved by CrabNet, whereas the top-two value of 8.11 corresponds to imKT@ModernBERT (Table S5). Excellent performance of this imKT model relates to advances in the BERT architecture (ModernBERT outperforms BERT and RoFormer, Tables S1–S5) and to the cross-modal knowledge transfer: multimodal learning as a pretraining task yields accuracy on par with the two-step procedure (Tables S6–S10). To sum up, the imKT technique ensures the highest in-class, i.e., across CLMs, performance (according to LLM4Mat-Bench results), and approaching accuracy of the best-performing architecture (i.e., CrabNet), as it follows from the MatBench results.

\begin{figure}
  \centering
  \includegraphics{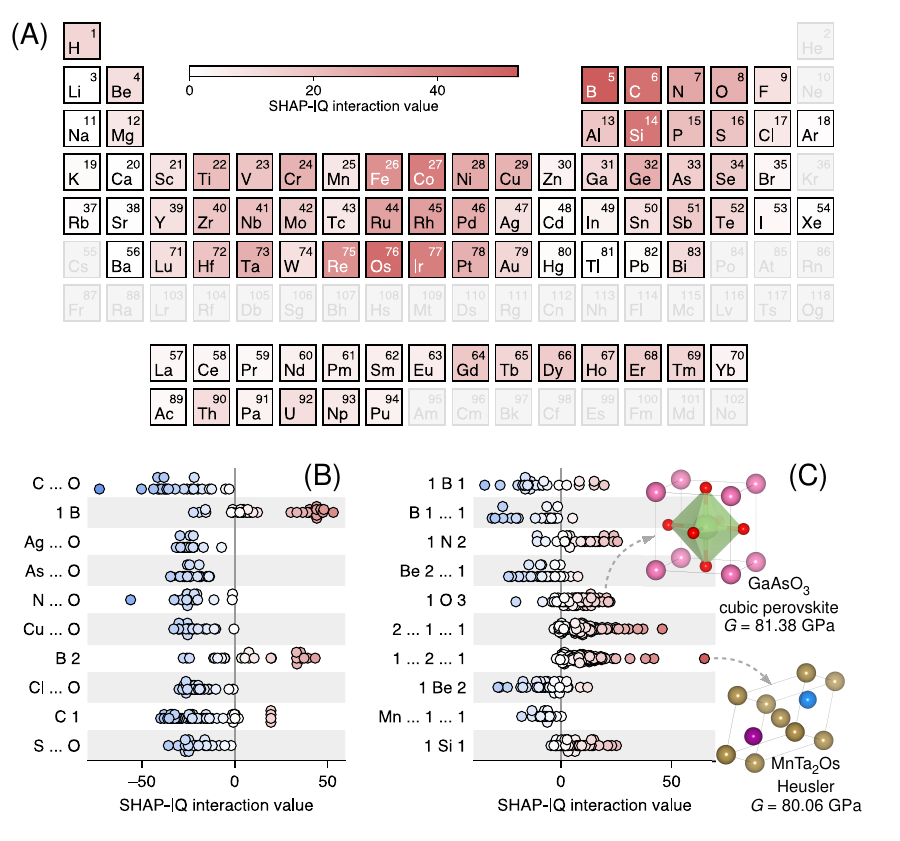}
  \caption{\textbf{Explainability analysis of chemical language model predicting shear modulus.} (A) Element-wise importance scores computed as averaged SHAPley Interaction Quantification (SHAP-IQ) values. Most influential (B) two- and (C) three-token combinations, according to the averaged SHAP-IQ values. Two crystal structures from the JARVIS-DFT dataset are depicted to outline most common structural prototypes; the corresponding DFT-computed shear moduli are provided as well.}
  \label{fig:fig2}
\end{figure}

Analyzing Tables S1–S5 reveals that the imKT models surpass the exKT models across all tasks, lowering MAE by 7.4\% on average. We assume that including structural information in the prediction pipeline does not improve accuracy because of two reasons. First, DFT-computed benchmarks contain a substantial fraction of metastable compounds, which are not a target of CSP. For instance, 37\% of the JARVIS-DFT compounds have an energy above the convex hull exceeding 0.1 eV/atom, and 6\% are highly unstable (${E}_{hull}$ $>$ 1.0 eV/atom). Second, the applied CSP approach has limited ability to output the most stable polymorphs: considering the JARVIS-DFT compounds with ${E}_{hull}$ $<$ 0.01 eV/atom, only a minor portion (11.7\%) of the generated crystal structures suites the original ones, in accordance with the structure matching procedure implemented in the pymatgen\cite{ong2013python}. Whereas the former issue is inherent to the task, the latter can be addressed by further developing high-throughput CSP algorithms. It should be also noted that training and inference of GPT-like and GNN models (used for generating crystal structures and reproducing target quantities, respectively) may result in substantial computational demands, which are another reason to prioritize CLMs as composition-based property predictors. Nevertheless, we consider exKT as a promising technique, taking into account its current performance: CLMs presented in LLM4Mat-Bench were outperformed in 22 out of 24 cases, summarizing the JARVIS-DFT and SNUMAT tasks (Tables S1–S3).

Neural networks for materials property prediction are typically utilized as “black-boxes”, missing understanding of their internal machinery. A few exceptions among composition-based models are transformers with attention attributes\cite{wang2021compositionally,ihalage2022formula}, which are debatably related to feature importances\cite{jain2019attention}. In this context, we applied a post hoc explainability technique that includes any-order feature interactions to gain insights into structure-property relationships, going beyond element-wise contributions. Specifically, the outputs of the imKT@ModernBERT model for predicting shear modulus (JARVIS-DFT dataset, LLM4Mat-Bench) were processed using the SHAPley Interaction Quantification\cite{muschalik2024shapiq} (SHAP-IQ) approach. As a starting point for analysis, the averaged elemental contributions are provided in Figure 2A. Expectedly, the highest impact on the shear modulus value is exerted by the presence of some platinum group metals and non-metals forming (ultra)hard materials, e.g., borides, carbides, and silicides. Considering the most influential two-token combinations (Figure 2B), we conclude that oxygen-containing groups reduce the target quantity. In contrast, boron-containing sequences increase the shear modulus value in most cases, though further insights can hardly be gained. Three-token combinations are most informative features, which extend our understanding beyond a compositional aspect (Figure 2C). In particular, the “\texttt{1 O 3}” sequence relates to the cubic perovskite prototype (CaTiO${}_{3}$, E${2}_{1}$ \textit{Strukturbericht} designation, $Pm\bar{3}m$ space group) in 38\% of cases, as it follows from analysis of the corresponding JARVIS-DFT crystal structures. The Heusler structure (AlCu${}_{2}$Mn, L${2}_{1}$ Strukturbericht designation, $Fm\bar{3}m$ space group) was identified for a noticeable part of “\texttt{2\ldots1\ldots 1}” and “\texttt{1\ldots2\ldots1}” sequences (41\% and 35\% of cases, respectively). However, there are no preferable prototypes for other three-token sequences. These findings represent an early example of interpreting chemical compositions through the crystallographic prototype analysis and advanced game-theoretic approach (i.e., SHAP-IQ), which is a promising combination for enhancing CLM explainability.

The presented approach—cross-modal knowledge transfer—is proved to be an effective route for enhancing efficacy of composition-based materials property prediction. Its modular structure (in both formulations) allows replacing distinct components. In the case of implicit transfer, we expect further enhancement owing to the development of multimodal representation learning and CLMs. On the other hand, explicit transfer can be advanced by resolving issues related to generating unplausible crystal structures; introducing multimodal foundational models as property predictors operating on hypothetical structures is another avenue for empowering the exKT pipeline. Finally, advanced ensembling techniques, e.g., mixture of experts, were shown to be a powerful auxiliary tool for increasing accuracy of heterogenous property predictors\cite{almeida2025multi}; two formulations of cross-modal knowledge transfer can be unified within this approach.

\section{Conflicts of interest}
\label{sec:conflicts}
There are no conflicts of interest to declare.

\section{Funding}
\label{sec:funding}
This work was supported by the Ministry of Economic Development of the Russian Federation in accordance with the subsidy agreement (agreement identifier 000000C313925P4H0002; grant No 139-15-2025-012).

\section{Acknowledgments}
\label{sec:acknowledgements}
We are grateful to Viggo Moro and other developers of the MultiMat framework for sharing model checkpoint, which was used for pretraining chemical language models.

\section{Data and software availability}
\label{sec:data_availability}
Source code for preprocessing data and training models can be found at \url{https://github.com/korolewadim/multimat-modernbert}. The ModernBERT-base model pretrained via a two-step procedure is available at \url{https://huggingface.co/korolewadim/multimat-modernbert}.

\bibliographystyle{unsrt}
\bibliography{references}

@article{zunger2018inverse,
  title={Inverse design in search of materials with target functionalities},
  author={Zunger, Alex},
  journal={Nature Reviews Chemistry},
  volume={2},
  number={4},
  pages={0121},
  year={2018},
  publisher={Nature Publishing Group UK London}
}

@article{butler2018machine,
  title={Machine learning for molecular and materials science},
  author={Butler, Keith T and Davies, Daniel W and Cartwright, Hugh and Isayev, Olexandr and Walsh, Aron},
  journal={Nature},
  volume={559},
  number={7715},
  pages={547--555},
  year={2018},
  publisher={Nature Publishing Group UK London}
}

@article{reiser2022graph,
  title={Graph neural networks for materials science and chemistry},
  author={Reiser, Patrick and Neubert, Marlen and Eberhard, Andr{\'e} and Torresi, Luca and Zhou, Chen and Shao, Chen and Metni, Houssam and van Hoesel, Clint and Schopmans, Henrik and Sommer, Timo and others},
  journal={Communications Materials},
  volume={3},
  number={1},
  pages={93},
  year={2022},
  publisher={Nature Publishing Group UK London}
}

@article{xie2018crystal,
  title={Crystal graph convolutional neural networks for an accurate and interpretable prediction of material properties},
  author={Xie, Tian and Grossman, Jeffrey C},
  journal={Physical review letters},
  volume={120},
  number={14},
  pages={145301},
  year={2018},
  publisher={APS}
}

@article{chen2019graph,
  title={Graph networks as a universal machine learning framework for molecules and crystals},
  author={Chen, Chi and Ye, Weike and Zuo, Yunxing and Zheng, Chen and Ong, Shyue Ping},
  journal={Chemistry of Materials},
  volume={31},
  number={9},
  pages={3564--3572},
  year={2019},
  publisher={ACS Publications}
}

@article{choudhary2021atomistic,
  title={Atomistic line graph neural network for improved materials property predictions},
  author={Choudhary, Kamal and DeCost, Brian},
  journal={npj Computational Materials},
  volume={7},
  number={1},
  pages={185},
  year={2021},
  publisher={Nature Publishing Group UK London}
}

@article{sole2025cartesian,
  title={A Cartesian encoding graph neural network for crystal structure property prediction: application to thermal ellipsoid estimation},
  author={Sol{\'e}, {\`A}lex and Mosella-Montoro, Albert and Cardona, Joan and G{\'o}mez-Coca, Silvia and Aravena, Daniel and Ruiz, Eliseo and Ruiz-Hidalgo, Javier},
  journal={Digital Discovery},
  volume={4},
  number={3},
  pages={694--710},
  year={2025},
  publisher={Royal Society of Chemistry}
}

@article{moro2025multimodal,
  title={Multimodal foundation models for material property prediction and discovery},
  author={Moro, Viggo and Loh, Charlotte and Dangovski, Rumen and Ghorashi, Ali and Ma, Andrew and Chen, Zhuo and Kim, Samuel and Lu, Peter Y and Christensen, Thomas and Solja{\v{c}}i{\'c}, Marin},
  journal={Newton},
  volume={1},
  number={1},
  year={2025},
  publisher={Elsevier}
}

@article{davies2016computational,
  title={Computational screening of all stoichiometric inorganic materials},
  author={Davies, Daniel W and Butler, Keith T and Jackson, Adam J and Morris, Andrew and Frost, Jarvist M and Skelton, Jonathan M and Walsh, Aron},
  journal={Chem},
  volume={1},
  number={4},
  pages={617--627},
  year={2016},
  publisher={Elsevier}
}

@article{oganov2019structure,
  title={Structure prediction drives materials discovery},
  author={Oganov, Artem R and Pickard, Chris J and Zhu, Qiang and Needs, Richard J},
  journal={Nature Reviews Materials},
  volume={4},
  number={5},
  pages={331--348},
  year={2019},
  publisher={Nature Publishing Group UK London}
}

@article{alghadeer2024machine,
  title={Machine learning prediction of materials properties from chemical composition: Status and prospects},
  author={Alghadeer, Mohammad and Aisyah, Nufida D and Hezam, Mahmoud and Alqahtani, Saad M and Baloch, Ahmer AB and Alharbi, Fahhad H},
  journal={Chemical Physics Reviews},
  volume={5},
  number={4},
  year={2024},
  publisher={AIP Publishing}
}

@article{ward2016general,
  title={A general-purpose machine learning framework for predicting properties of inorganic materials},
  author={Ward, Logan and Agrawal, Ankit and Choudhary, Alok and Wolverton, Christopher},
  journal={npj Computational Materials},
  volume={2},
  number={1},
  pages={1--7},
  year={2016},
  publisher={Nature Publishing Group}
}

@article{zhuo2018predicting,
  title={Predicting the band gaps of inorganic solids by machine learning},
  author={Zhuo, Ya and Mansouri Tehrani, Aria and Brgoch, Jakoah},
  journal={The journal of physical chemistry letters},
  volume={9},
  number={7},
  pages={1668--1673},
  year={2018},
  publisher={ACS Publications}
}

@article{ghiringhelli2015big,
  title={Big data of materials science: critical role of the descriptor},
  author={Ghiringhelli, Luca M and Vybiral, Jan and Levchenko, Sergey V and Draxl, Claudia and Scheffler, Matthias},
  journal={Physical review letters},
  volume={114},
  number={10},
  pages={105503},
  year={2015},
  publisher={APS}
}

@article{ouyang2018sisso,
  title={SISSO: A compressed-sensing method for identifying the best low-dimensional descriptor in an immensity of offered candidates},
  author={Ouyang, Runhai and Curtarolo, Stefano and Ahmetcik, Emre and Scheffler, Matthias and Ghiringhelli, Luca M},
  journal={Physical Review Materials},
  volume={2},
  number={8},
  pages={083802},
  year={2018},
  publisher={APS}
}

@article{jha2018elemnet,
  title={Elemnet: Deep learning the chemistry of materials from only elemental composition},
  author={Jha, Dipendra and Ward, Logan and Paul, Arindam and Liao, Wei-keng and Choudhary, Alok and Wolverton, Chris and Agrawal, Ankit},
  journal={Scientific reports},
  volume={8},
  number={1},
  pages={17593},
  year={2018},
  publisher={Nature Publishing Group UK London}
}

@article{goodall2020predicting,
  title={Predicting materials properties without crystal structure: deep representation learning from stoichiometry},
  author={Goodall, Rhys EA and Lee, Alpha A},
  journal={Nature communications},
  volume={11},
  number={1},
  pages={6280},
  year={2020},
  publisher={Nature Publishing Group UK London}
}

@article{wang2021compositionally,
  title={Compositionally restricted attention-based network for materials property predictions},
  author={Wang, Anthony Yu-Tung and Kauwe, Steven K and Murdock, Ryan J and Sparks, Taylor D},
  journal={Npj Computational Materials},
  volume={7},
  number={1},
  pages={77},
  year={2021},
  publisher={Nature Publishing Group UK London}
}

@article{chen2021atomsets,
  title={AtomSets as a hierarchical transfer learning framework for small and large materials datasets},
  author={Chen, Chi and Ong, Shyue Ping},
  journal={npj Computational Materials},
  volume={7},
  number={1},
  pages={173},
  year={2021},
  publisher={Nature Publishing Group UK London}
}

@article{ihalage2022formula,
  title={Formula Graph Self-Attention Network for Representation-Domain Independent Materials Discovery},
  author={Ihalage, Achintha and Hao, Yang},
  journal={Advanced Science},
  volume={9},
  number={18},
  pages={2200164},
  year={2022},
  publisher={Wiley Online Library}
}

@article{tshitoyan2019unsupervised,
  title={Unsupervised word embeddings capture latent knowledge from materials science literature},
  author={Tshitoyan, Vahe and Dagdelen, John and Weston, Leigh and Dunn, Alexander and Rong, Ziqin and Kononova, Olga and Persson, Kristin A and Ceder, Gerbrand and Jain, Anubhav},
  journal={Nature},
  volume={571},
  number={7763},
  pages={95--98},
  year={2019},
  publisher={Nature Publishing Group}
}

@article{vaswani2017attention,
  title={Attention is all you need},
  author={Vaswani, Ashish and Shazeer, Noam and Parmar, Niki and Uszkoreit, Jakob and Jones, Llion and Gomez, Aidan N and Kaiser, {\L}ukasz and Polosukhin, Illia},
  journal={Advances in neural information processing systems},
  volume={30},
  year={2017}
}

@inproceedings{velivckovic2018graph,
  title={Graph Attention Networks},
  author={Veli{\v{c}}kovi{\'c}, Petar and Cucurull, Guillem and Casanova, Arantxa and Romero, Adriana and Li{\`o}, Pietro and Bengio, Yoshua},
  booktitle={International Conference on Learning Representations},
  year={2018}
}

@article{na2025cross,
  title={Cross-modality material embedding loss for transferring knowledge between heterogeneous material descriptors},
  author={Na, Gyoung S},
  journal={npj Computational Materials},
  volume={11},
  number={1},
  pages={235},
  year={2025},
  publisher={Nature Publishing Group UK London}
}

@article{trewartha2022quantifying,
  title={Quantifying the advantage of domain-specific pre-training on named entity recognition tasks in materials science},
  author={Trewartha, Amalie and Walker, Nicholas and Huo, Haoyan and Lee, Sanghoon and Cruse, Kevin and Dagdelen, John and Dunn, Alexander and Persson, Kristin A and Ceder, Gerbrand and Jain, Anubhav},
  journal={Patterns},
  volume={3},
  number={4},
  year={2022},
  publisher={Elsevier}
}

@article{niyongabo2025llm,
  title={LLM-Prop: predicting the properties of crystalline materials using large language models},
  author={Niyongabo Rubungo, Andre and Arnold, Craig and Rand, Barry P and Dieng, Adji Bousso},
  journal={npj Computational Materials},
  volume={11},
  number={1},
  pages={186},
  year={2025},
  publisher={Nature Publishing Group UK London}
}

@article{rubungo2025llm4mat,
  title={LLM4Mat-bench: benchmarking large language models for materials property prediction},
  author={Rubungo, Andre Niyongabo and Li, Kangming and Hattrick-Simpers, Jason and Dieng, Adji Bousso},
  journal={Machine Learning: Science and Technology},
  volume={6},
  number={2},
  pages={020501},
  year={2025},
  publisher={IOP Publishing}
}

@article{antunes2024crystal,
  title={Crystal structure generation with autoregressive large language modeling},
  author={Antunes, Luis M and Butler, Keith T and Grau-Crespo, Ricardo},
  journal={Nature Communications},
  volume={15},
  number={1},
  pages={10570},
  year={2024},
  publisher={Nature Publishing Group UK London}
}

@article{ong2013python,
  title={Python Materials Genomics (pymatgen): A robust, open-source python library for materials analysis},
  author={Ong, Shyue Ping and Richards, William Davidson and Jain, Anubhav and Hautier, Geoffroy and Kocher, Michael and Cholia, Shreyas and Gunter, Dan and Chevrier, Vincent L and Persson, Kristin A and Ceder, Gerbrand},
  journal={Computational Materials Science},
  volume={68},
  pages={314--319},
  year={2013},
  publisher={Elsevier}
}

@article{muschalik2024shapiq,
  title={shapiq: Shapley interactions for machine learning},
  author={Muschalik, Maximilian and Baniecki, Hubert and Fumagalli, Fabian and Kolpaczki, Patrick and Hammer, Barbara and H{\"u}llermeier, Eyke},
  journal={Advances in Neural Information Processing Systems},
  volume={37},
  pages={130324--130357},
  year={2024}
}

@inproceedings{jain2019attention,
  title={Attention is not Explanation},
  author={Jain, Sarthak and Wallace, Byron C},
  booktitle={Proceedings of the 2019 Conference of the North American Chapter of the Association for Computational Linguistics: Human Language Technologies, Volume 1 (Long and Short Papers)},
  pages={3543--3556},
  year={2019}
}

@article{almeida2025multi,
  title={Multi-View Mixture-of-Experts for predicting molecular properties using SMILES, SELFIES, and graph-based representations},
  author={Almeida Soares, Eduardo and Shirasuna, Victor and Vital Brazil, Emilio and Priyadarsini S, Indra and Takeda, Seiji},
  journal={Machine Learning: Science and Technology},
  year={2025},
  pages={025070},
  number={2},
  volume={6},
}

@article{dunn2020benchmarking,
  title={Benchmarking materials property prediction methods: the Matbench test set and Automatminer reference algorithm},
  author={Dunn, Alexander and Wang, Qi and Ganose, Alex and Dopp, Daniel and Jain, Anubhav},
  journal={npj Computational Materials},
  volume={6},
  number={1},
  pages={138},
  year={2020},
  publisher={Nature Publishing Group UK London}
}

\end{document}